\documentstyle[preprint,aps,epsf,floats,tighten]{revtex} 

\begin{document}

\preprint{\tighten \vbox{\hbox{}
\hbox{} \hbox{} \hbox{} \hbox{} \hbox{} } }

\title{Determinations of $|V_{ub}|$ and $|V_{cb}|$ from Measurements of
$B\to X_{u,c}\ell\nu$ Differential Decay Rates}

\author{Changhao Jin}

\address{School of Physics,
University of Melbourne\\Parkville, Victoria 3052, Australia}

\maketitle

{\tighten
\begin{abstract}%
Methods are described in the framework of light-cone expansion which allow
one to determine the Cabibbo-Kobayashi-Maskawa matrix 
elements $|V_{ub}|$ and $|V_{cb}|$ from measurements of the differential decay
rates as a function of the scaling variables in the inclusive semileptonic 
decays of $B$ mesons. By these model-independent methods the dominant hadronic
uncertainties can be avoided and the $B\to X_u\ell\nu$ decay can be very
efficiently differentiated from the 
$B\to X_c\ell\nu$ decay, which may lead to precise determinations of $|V_{ub}|$
and $|V_{cb}|$.
\end{abstract}
}

\newpage

The origins of quark masses, quark flavor mixing and CP violation are among 
the fundamental issues of particle physics. In the standard model, quark 
flavor mixing is described by the Cabibbo-Kobayashi-Maskawa (CKM) matrix 
\cite{ckm} and the phase of it is also responsible for the CP 
violation. The CKM matrix appears to be deeply connected to the origin and 
the values of quark masses,
since it results from the diagonalization of the quark mass 
matrices \cite{mass}.  The CKM matrix is unpredictable except for unitarity
in the standard model, and precise 
determinations of the magnitudes and the phases of its elements and a 
consistency checking in the results are of particular importance. 

The magnitudes of two CKM matrix elements $|V_{ub}|$ and $|V_{cb}|$ can be
determined from inclusive and exclusive semileptonic $B$ meson 
decays \cite{rev}. 
Main theoretical uncertainties in these
determinations arise from hadronic bound state effects on the underlying
weak decays. Precise determinations of $|V_{ub}|$ and $|V_{cb}|$ require
methods which reduce or avoid hadronic uncertainties.
Especially, improved precision in $|V_{ub}|$ is an even more pressing need.

The light-cone expansion provides a solution to the problem of organizing 
nonperturbative QCD effects on inclusive semileptonic decays of the $B$ meson
in such a way as to exploit heaviness of the $B$
meson \cite{jp,jin1,jp1,jin2} and allows the selection of observables less
affected by hadronization and thus better suited for precise determinations of
$|V_{ub}|$ and $|V_{cb}|$. In this framework the leading nonperturbative
effect is described by the distribution function of the $b$ quark inside
the $B$ meson. Several important properties of the distribution function
are known in QCD. In particular, the distribution function is normalized 
exactly to unity as a consequence of $b$ quantum number conservation.

In this paper, we propose to determine $|V_{ub}|$ and $|V_{cb}|$ by measuring
the differential decay rates as a function of 
$\xi_q=(\nu+\sqrt{\nu^2-q^2+m_q^2})/M$ in the inclusive semileptonic $B$ meson
decays $B\to X_q\ell\nu$, where the invariant variable $\nu$ is defined by 
$\nu=q\cdot P/M$, $q$ stands for the momentum transfer to the lepton pair,
$P$ and $M$ represent, respectively, the four-momentum and the mass of the 
$B$ meson, and $m_q$ is the final quark mass with $q=u, c$.
We will show that the differential decay rates $d\Gamma/d\xi_q$ are explicitly
proportional to the distribution function. Thus the dominant hadronic 
uncertainties can be avoided with these observables 
either by exploiting the known normalization of the
distribution function or through the cancellation of the distribution function
in the ratio of the differential decay rates.
These methods are complementary to other methods.  The main advantage of our
approach is that the dominant hadronic uncertainties can be avoided, providing
model-independent and precise determinations of $|V_{ub}|$ and $|V_{cb}|$, and 
the $B\to X_u\ell\nu$ decay can be more efficiently
differentiated from the $B\to X_c\ell\nu$ decay for determining $|V_{ub}|$ than
the proposed method to measure $|V_{ub}|$ by the hadronic invariant mass 
spectrum \cite{parton,barger,afalk,kim,bdu,jin3}.

The decay rate for the inclusive semileptonic $B$ meson decay
is given by
\begin{equation}
d\Gamma= \frac{G_{F}^2\left|V_{qb}\right|^2}{(2\pi)^5E}L^{\mu\nu}
W_{\mu\nu}\frac{d^3k_\ell}{2E_{\ell}}\frac{d^3k_{\nu}}{2E_{\nu}} ,
\label{eq:kga}
\end{equation}
where $E$ is the energy of the decaying $B$ meson, $k_{\ell(\nu)}$ and
$E_{\ell(\nu)}$ denote the four-momentum and the energy of the charged lepton 
(antineutrino), respectively. $V_{qb}$ is the element of the 
CKM matrix, which is $V_{cb}$ for the $b\to c$
transition induced decay and $V_{ub}$ for the $b\to u$ transition induced
decay. Our discussion is made in arbitrary frame of reference throughout this
work.
$L_{\mu\nu}$ is the leptonic tensor:
\begin{equation}
L^{\mu\nu}= 2(k^{\mu}_{\ell}k^{\nu}_{\nu}+k^{\mu}_{\nu}k^{\nu}_{\ell}-
g^{\mu\nu}k_{\ell}\cdot k_{\nu}+i\varepsilon^{\mu\nu}\hspace{0.06cm}_{\alpha\beta}
k^{\alpha}_{\ell}k^{\beta}_{\nu}).
\label{eq:klepton}
\end{equation}
$W_{\mu\nu}$ is the hadronic tensor:
\begin{equation}
W_{\mu\nu}= -\frac{1}{2\pi}\int d^4y\, e^{iq\cdot y}
\langle B\left|[j_{\mu}(y),j^{\dagger}_{\nu}(0)]\right|B\rangle ,
\label{eq:comm2}
\end{equation}
where $j_{\mu}(y) = \bar{q}(y)\gamma_{\mu}(1-\gamma_5)b(y)$ is the weak 
current.
The $B$ meson state $|B\rangle$ satisfies the standard covariant
normalization 
$\langle B|B\rangle=2E(2\pi)^3\delta^3({\bf 0})$.  

The nonperturbative QCD effects on the processes reside in the hadronic
tensor. In general, the hadronic tensor can be constructed of scalar structure 
functions by Lorentz decomposition, and three structure functions 
$W_a(\nu, \,q^2)$, $a = 1, 2, 3$ contribute
to the inclusive semileptonic $B$ decays under consideration with negligible
lepton masses: 
\begin{equation}
W_{\mu\nu} = -g_{\mu\nu}W_1 + \frac{P_{\mu}P_{\nu}}{M^2} W_2 
 -i\varepsilon_{\mu\nu\alpha\beta} \frac{P^{\alpha}q^{\beta}}{M^2}W_3\, .
\label{eq:exp2}
\end{equation}
    
The inclusive semileptonic decay of the $B$ meson involves large momentum
transfer over most of the phase space because of heaviness of 
the decaying $B$ meson.   Therefore,
the integral of Eq.~(\ref{eq:comm2}) is dominated by the light-cone
distance in the space-time structure.   
This leads to a factorization of the matrix element in Eq.~(\ref{eq:comm2}) 
into two parts: one characterizing the light-cone space-time singularity and
another being the reduced matrix element of the bilocal $b$ quark 
operator \cite{jp,jin1,jp1,jin2}:
\begin{equation}
\langle B| \left[ j_{\mu}(y),j_{\nu}^{\dagger}(0) \right] |B\rangle
 = 2(S_{\mu\alpha\nu\beta} -i\varepsilon_{\mu\alpha\nu\beta})
  \left[ \partial^{\alpha}\Delta_q(y) \right] \langle B|\bar{b}(0)
    \gamma^{\beta}(1-\gamma_5)b(y)|B\rangle\, ,
\label{eq:domin3}
\end{equation}
where $S_{\mu\alpha\nu\beta} = g_{\mu\alpha}g_{\nu\beta} + g_{\mu\beta}
 g_{\nu\alpha} - g_{\mu\nu}g_{\alpha\beta}$ 
and $\Delta_q(y)$ is the Pauli--Jordan function for a free $q$-quark of mass 
$m_q$. The light-cone expansion of the reduced matrix element provides a 
systematic way to organize the nonperturbative effects. The leading term of 
this expansion is given by
\begin{equation}
\langle B|\bar{b}(0)\gamma^\beta(1-\gamma_5)b(y)|B\rangle_{y^2=0}
= 2P^\beta\int_0^1 d\xi\, e^{-i\xi y\cdot P}f(\xi) .
\label{eq:lead}
\end{equation}
The function $f(\xi)$ is known as the distribution function of the $b$ quark 
inside the $B$ meson.
A similar distribution function has been introduced by the resummation of the
heavy quark expansion \cite{resum}.

The light-cone dominance implies that the structure functions are given 
by \cite{jp,jin1,jp1}
\begin{eqnarray}
W_1 & = & 2[f(\xi_+) + f(\xi_-)]\, ,\label{eq:w1}\\
W_2 & = & \frac{8}{\xi_+ -\xi_-}[\xi_+f(\xi_+)-\xi_-f(\xi_-)]\,,\label{eq:w2}\\
W_3 & = & -\frac{4}{\xi_+-\xi_-} [f(\xi_+) -f(\xi_-)]\, ,\label{eq:w3}
\end{eqnarray}
where
\begin{equation}
\xi_{\pm}=\frac{\nu\pm\sqrt{\nu^2-q^2+m_q^2}}{M}\, .
\label{eq:root}
\end{equation} 
Equations (\ref{eq:w1})--(\ref{eq:w3}) include the leading twist contribution;
higher-twist contributions are expected to be suppressed by powers of $q^2$.

The leading nonperturbative QCD effects on the inclusive semileptonic decays 
is now encoded in the distribution function $f(\xi)$. 
The detailed form of the distribution function is not known.
However, taking $y=0$ in Eq.(\ref{eq:lead}), $b$ quantum number conservation 
implies the normalization 
condition \cite{jp,jin2}:
\begin{equation}
\int_0^1 d\xi f(\xi)=\frac{1}{2M^2}P^\mu\langle B\left|\bar b(0)
\gamma_\mu
(1-\gamma_5)b(0)\right| B\rangle = 1 .
\label{eq:norm}
\end{equation}
  
The differential decay rate as a function of the scaling variable $\xi_+$
for $B\to X_q\ell\nu$ is calculated in terms of the distribution function as
\begin{eqnarray}
\frac{d\Gamma}{d\xi_+} =&&\frac{G_F^2|V_{qb}|^2}{48\pi^3}\,
  \frac{M}{E}\Bigg\{ \frac{1}{4}M^5\xi_+^5f(\xi_+)\Phi(r_q/\xi_+) \nonumber\\
&&+\int_0^{(\xi_+M-m_q)^2}dq^2\,\sqrt{\nu^2-q^2}f(\xi_-)\Bigg [3q^2+
\frac{\xi_-}{\xi_+}(q^2-4\nu^2)\Bigg ]\Bigg\} \, ,
\label{eq:rate1}
\end{eqnarray}
with
\begin{equation}
\Phi(x)=1-8x^2+8x^6-x^8-24x^4\mbox{ln}x \, ,
\end{equation}
where $r_q=m_q/M$, $\nu$ and $\xi_-$ are related to $\xi_+$ and $q^2$ by
\begin{eqnarray}
\nu & = & \frac{\xi_+^2M^2+q^2-m_q^2}{2\xi_+M} \, ,\\
\xi_- & = & \frac{q^2-m_q^2}{\xi_+M^2} \, .
\end{eqnarray}
The kinematic range of $\xi_+$ is $r_q\leq\xi_+\leq 1$. Note that $\xi_+$ is
different kinematic variable for $B\to X_u\ell\nu$ and $B\to X_c\ell\nu$, 
defined in Eq.~(\ref{eq:root}).

The free quark limit corresponds to $f(\xi)=\delta (\xi-m_b/M)$.
Since the heavy $b$ quark inside the $B$ meson is almost on-shell,
we expect that the distribution function in reality is very close to the delta
function, which is supported by
the analysis based on the operator product expansion and the heavy quark 
effective theory (HQET) \cite{hqet,hqe}, indicating 
that the distribution function peaks sharply around 
$\xi=m_b/M\approx 0.93$ \cite{jp,jin1}.

The $f(\xi_-)$ term is a consequence of field theory, corresponding to
the creation of quark-antiquark pairs through the Z-graph.
The contribution of the $f(\xi_-)$ term, present as an integral in 
Eq.~(\ref{eq:rate1}), is expected to play less of a role, since (i) in the free
quark limit the integral vanishes and (ii) the dominant contribution to the
$f(\xi_-)$ integral at a given $\xi_+$ resulting from the large $\xi_-$ region,
corresponding to the neighbourhood of the upper integration limit for $q^2$,
is suppressed by $\nu^2-q^2$. This expectation is confirmed by the numerical
study discussed below.
The spectra $d\Gamma/d\xi_+$ are dominated by the $f(\xi_+)$ term for both
$b\to c$ and $b\to u$ transitions and both spectra have a sharp peak at the 
same peak location as $f(\xi_+)$. In the following we will to good 
approximation ignore the $f(\xi_-)$ term and so the differential 
decay rates $d\Gamma/d\xi_+$ given by Eq.~(\ref{eq:rate1}) are proportional 
to $f(\xi_+)$.

In the free quark limit, the tree-level and 
virtual gluon processes $b\to u(c)\ell\nu$ generate a spectrum nonvanishing at
only one point, i.e., $\xi_+=m_b/M$, solely on kinematic grounds. It is gluon 
bremsstrahlung and hadronic bound state effects that expand the spectrum over 
the whole range of $\xi_+$ from $m_q/M$ to 1. This unique feature implies that
measurements of the spectra $d\Gamma/d\xi_+$ would offer the intrinsically 
most sensitive probe of long-distance strong interactions. Indeed, we have 
shown on the basis of light-cone expansion that the spectra $d\Gamma/d\xi_+$ 
are explicitly proportional to the nonperturbative distribution function.
Thus their measurements would lead to a direct extraction of the distribution 
function (see further below). On the other hand, the spectra $d\Gamma/d\xi_+$
also provide the most straightforward and best way to eliminate the dependence
on the distribution function since they are proportional to it, so that the
dominant hadronic uncertainties can be avoided. Moreover, the 
$B\to X_q\ell\nu$ spectra $d\Gamma/d\xi_+$ are sharply peaked at 
$\xi_+= m_b/M$, since the spectra stemming from the tree-level and virtual 
gluon processes would only concentrate at $\xi_+= m_b/M$ and gluon 
bremsstrahlung and hadronic bound state effects smear the spectra about this 
point, but most of the decay rates remain around $\xi_+= m_b/M$. This implies
that the kinematic cut on the $b\to u$ scaling variable $\xi_+$ will make a
very efficient discrimination between $B\to X_u\ell\nu$ and $B\to X_c\ell\nu$
events, even better than the kinematic cut on the hadronic invariant mass of
the final state.
Consequently, measurements of the decay distribution as a function of the 
$b\to u$ scaling variable $\xi_+$ would yield, particularly, a high precision 
$|V_{ub}|$ determination. We will discuss these points in detail below. 

For the charmless decay $B\to X_u\ell\nu$, we obtain from 
Eq.~(\ref{eq:rate1})
\begin{equation}
|V_{ub}|^2f(\xi_u)=\frac{192\pi^3E}{G_F^2M^6}\frac{1}{\xi_u^5}
\frac{d\Gamma(B\to X_u\ell\nu)}{d\xi_u}\, ,
\label{eq:bu}
\end{equation}
with $\xi_u=(\nu_u+\sqrt{\nu_u^2-q_u^2})/M$. We have ignored the up quark mass.
To avoid confusion,  quantities for the $b\to u$ ($b\to c$) decay are 
indicated by a subscript $u$ ($c$); in particular, $\xi_+$ is explicitly 
denoted as $\xi_{u(c)}$ for the $b\to u$ ($b\to c$) decay.  

By integrating Eq.~(\ref{eq:bu}) over $\xi_u$ and using the normalization 
condition (\ref{eq:norm}),
one gets rid of the distribution function, obtaining
\begin{equation}
|V_{ub}|^2=\frac{192\pi^3E}{G_F^2M^6}\int_0^1d\xi_u\, \frac{1}{\xi_u^5}
\frac{d\Gamma(B\to X_u\ell\nu)}{d\xi_u}\, .
\label{eq:vub}
\end{equation}
This implies that the known normalization of the distribution function allows
a model independent determination of $|V_{ub}|$ from a measurement of the
weighted integral of the decay spectrum $d\Gamma(B\to X_u\ell\nu)/d\xi_u$. 

Likewise, for the charmed decay $B\to X_c\ell\nu$, we obtain from 
Eq.~(\ref{eq:rate1})
\begin{equation}
|V_{cb}|^2f(\xi_c)=\frac{192\pi^3E}{G_F^2M^6}\frac{1}{\xi_c^5\Phi(r_c/\xi_c)}
\frac{d\Gamma(B\to X_c\ell\nu)}{d\xi_c}\, ,
\label{eq:bc}
\end{equation}
with $\xi_c=(\nu_c+\sqrt{\nu_c^2-q_c^2+m_c^2})/M$.
Integrating Eq.~(\ref{eq:bc}) over $\xi_c$ and using the normalization 
condition (\ref{eq:norm}) yields
\begin{equation}
|V_{cb}|^2=\frac{192\pi^3E}{G_F^2M^6}\int_{r_c}^1d\xi_c\,
\frac{1}{\xi_c^5\Phi(r_c/\xi_c)}\frac{d\Gamma(B\to X_c\ell\nu)}
{d\xi_c}\, ,
\label{eq:vcb}
\end{equation}
which is independent of the distribution function. Thus, a measurement of the 
weighted integral of the charmed differential decay rate as a function
of $\xi_c$ would provide a model independent determination of $|V_{cb}|$.  

The distribution function cancels in the ratio of the differential decay   
rates at $\xi_u=\xi_c$ and we obtain from Eqs.~(\ref{eq:bu}) and
(\ref{eq:bc})
\begin{equation}
\Bigg |\frac{V_{ub}}{V_{cb}}\Bigg |^2=\Phi\Bigg (\frac{r_c}{\xi_c}\Bigg )
\Bigg [\frac{d\Gamma(B
\to X_u\ell\nu)/d\xi_u}{d\Gamma(B\to X_c\ell\nu)/d\xi_c}
\Bigg ]_{\xi_u=\xi_c}\,\, .
\label{eq:ratio}
\end{equation}
This provides a model independent method for determining the ratio
$|V_{ub}/V_{cb}|$. 

Equation (\ref{eq:bc}) shows that a measurement of the decay spectrum
$d\Gamma(B\to X_c\ell\nu)/d\xi_c$ can also be used to extract the
distribution function directly.  We note that
the distribution function can also be extracted from a measurement 
of the charmless decay distribution with respect to $\xi_u$, as shown by
Eq.~(\ref{eq:bu}),
though the decay rate is much smaller than the $b\to c$ decay rate.

\begin{figure}[t]
\centerline{\epsfysize=9truecm \epsfbox{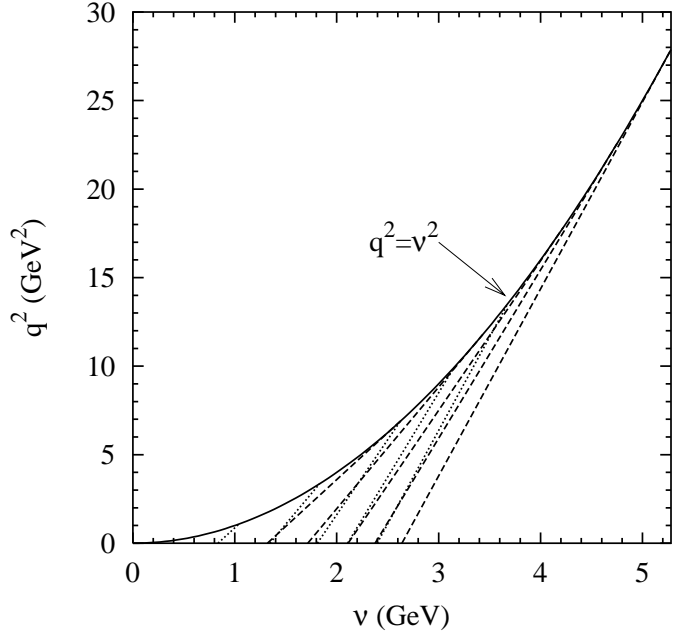}}
\tighten{
\caption[tau1]{Contours for $\xi_{u,c}$ in the ($\nu, q^2$) phase space. 
The dashed
(dotted) lines from right to left refer to $\xi_{u(c)}=1, 0.9, 0.8, 0.65,
0.5$, taking $m_c=1.6$ GeV. } 
\label{fig:contour} }
\end{figure}

The quantity $\xi_q (q=u, c)$ is a combination of two kinematic variables 
$\nu$ and $q^2$,  which can be
measured through neutrino reconstruction.  Except
the lower end point $\xi_q=r_q$ corresponding to $\nu=0$ and $q^2=0$,
different values of $\nu$ and $q^2$ correspond to the same value of $\xi_q$, 
as shown in Fig.~1,
thereby giving the same physical value for the differential decay rate 
$d\Gamma(B\to X_q\ell\nu)/d\xi_q$. This implies that the spectrum can 
be well described by the light-cone dynamics as long as it can originate 
from processes with sufficiently large momentum transfer.  
From Fig.~1,  we see that this 
is the case for the dominant spectrum over not too small values of
$\xi_q$.  Thus departures from the light cone are expected to 
bring little theoretical uncertainty in the methods previously described. 
Experimentally, one can take advantage of the multivariate freedom in   
$\xi_q$ to choose specific values of $\nu$ and $q^2$ in phase space 
so as to discriminate between $b\to u$ and $b\to c$ events with the use of 
kinematic cuts and eliminate the charm quark mass uncertainty. 
If one measures two different pairs of $(\nu_1, q^2_1)$ and $(\nu_2, q^2_2)$
giving the same decay rate $d\Gamma(B\to X_c\ell\nu)/d\xi_c$, this could mean
$\nu_1+\sqrt{\nu^2_1-q^2_1+m_c^2}= \nu_2+\sqrt{\nu^2_2-q^2_2+m^2_c}$, which
determines $m_c$ from experiment, thereby eliminating the charm quark mass
uncertainty.
Note that, as shown in Fig.~1, only the $b\to u$ transition
is kinematically allowed for $\xi_c>1$. This cut gives access to a larger 
portion of phase space than the cut directly on $\nu$ or $q^2$. 
The $b\to u$ and $b\to c$ events 
with $\xi_u=\xi_c$ must correspond to the different points in phase space,
as can be seen in Fig.~1, which renders the discrimination
between $b\to u$ and $b\to c$ events easy, in particular when using the 
method of Eq.~(\ref{eq:ratio}). 

To verify that the contributions of the $f(\xi_-)$ term in Eq.~(\ref{eq:rate1})
are negligible,
we have made a numerical study by using the HQET-constrained 
parametrization of \cite{jin1} for the distribution function. 
As expected, the contributions of
the $f(\xi_-)$ term to the integrated decay rates in Eqs.~(\ref{eq:vub}) and
(\ref{eq:vcb}) are both at the level below $1\%$, yielding 
negligible theoretical uncertainties in $|V_{ub}|$ and $|V_{cb}|$ obtained by
these two methods.

\begin{figure}[t]
\centerline{\epsfysize=8truecm \epsfbox{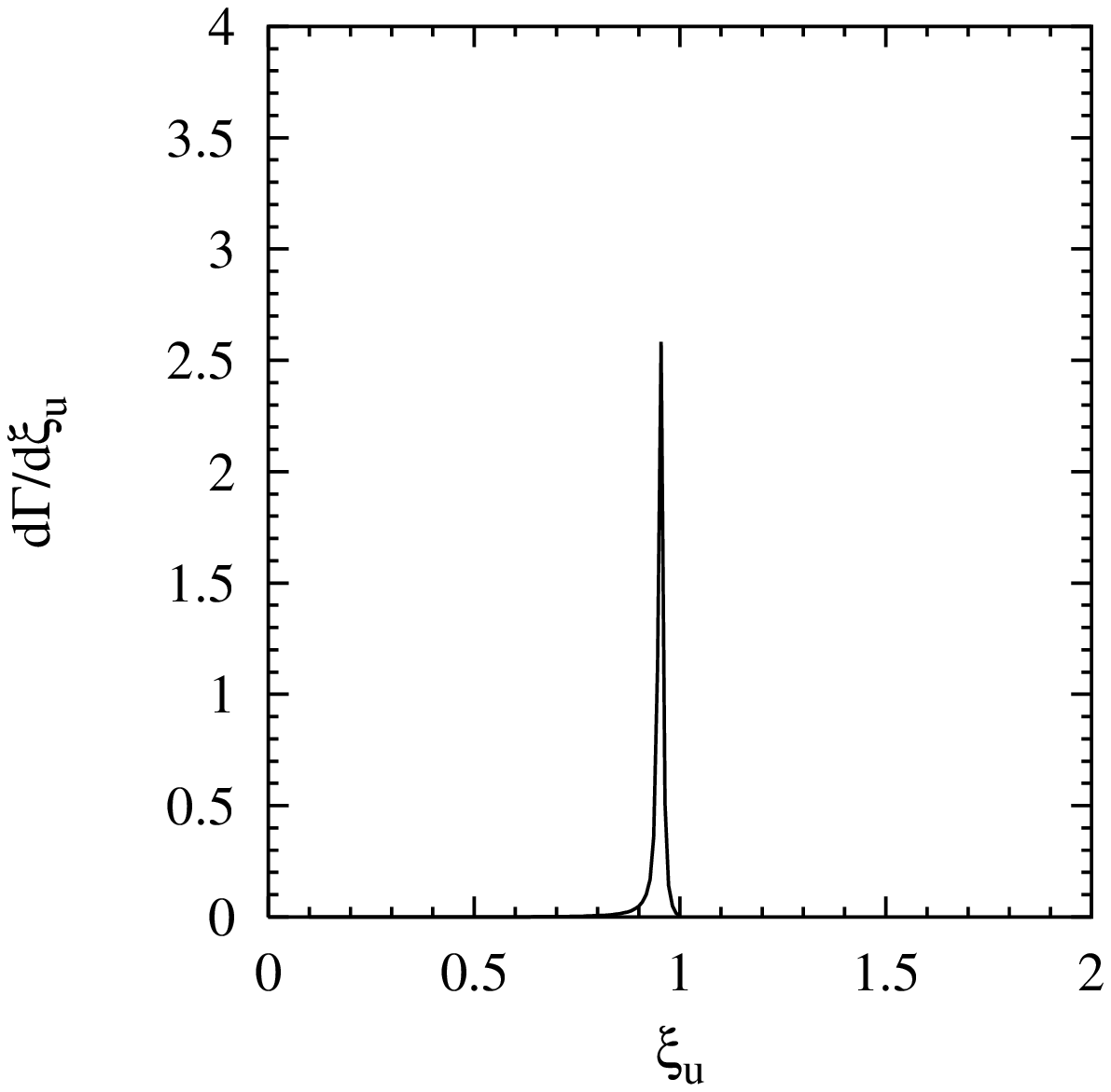}
  \epsfysize=8truecm \epsfbox{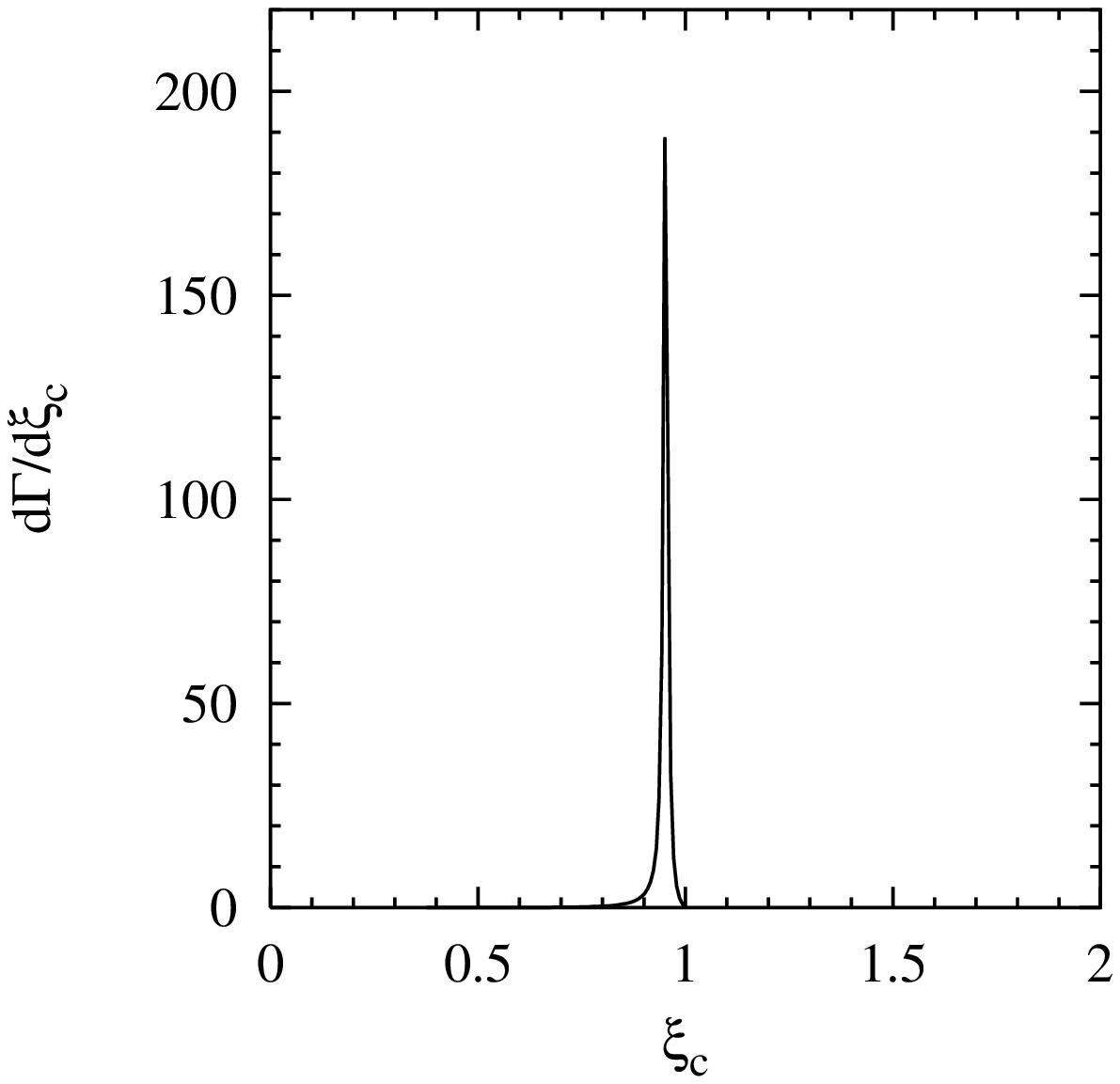}}
\tighten{
\caption[tau1]{The spectra $d\Gamma(B\to X_u\ell\nu)/d\xi_u$ (left) 
and $d\Gamma(B\to X_c\ell\nu)/d\xi_c$ (right) calculated using the
parametrization of \cite{jin1} for the distribution function for 
$\alpha=\beta=1$, $a=0.953, b=0.00560$, and $m_c=1.6$ GeV. The radiative QCD
corrections are not included. 
The absolute scale is arbitrary, but the relative scale between the $b\to u$
and $b\to c$ spectra is fixed to $|V_{ub}/V_{cb}|=0.08$. } 
\label{fig:spec} }
\end{figure}

Figure 2 shows the spectra 
$d\Gamma(B\to X_u\ell\nu)/d\xi_u$ and 
$d\Gamma(B\to X_c\ell\nu)/d\xi_c$. 
We see that both spectra peak sharply at the same value for $\xi_u$ and 
$\xi_c$.
This feature does not depend on the detailed form of the distribution function.
However, the shapes of both spectra are sensitive to the form of the 
distribution function.  Numerical studies show that
including the $f(\xi_-)$ term makes almost no
difference in the spectra, and the tiny
contribution from the $f(\xi_-)$ term is meaningful (although still invisible 
in Fig.~2)
only for the spectra very close to their two end points, where the spectra 
become vanishingly small.   
Therefore, measurements of both $\xi_u$ and $\xi_c$ distributions in the 
neighborhood of maximum, where the $f(\xi_-)$ term is negligible, 
would be appropriate for a reliable determination of 
$|V_{ub}/V_{cb}|$ by the method of Eq.~(\ref{eq:ratio}).
The smallness of the $f(\xi_-)$ term has already been found in the
parton model \cite{parton}. 

The determination of $|V_{cb}|$ by the method proposed in this paper may still
suffer from large theoretical systematic errors. The light-cone expansion 
parameter is $\Lambda^2_{\rm QCD}/q^2$. For $B\to X_c\ell\nu$, the maximum
value of $q^2$ is $(M-M_D)^2$ with $M_D$ being the $D$ meson mass. It means
that $q^2$ is not large enough to neglect the higher order corrections. The
$\Lambda^2_{\rm QCD}/q^2$ correction may be necessary to achieve $5\%$ 
accuracy 
for $|V_{cb}|$. Actually the semileptonic $b\to c$ decay rate is dominated by
a few exclusive decay modes ($D, D^\ast$ and $D^{(\ast)}\pi$), which suggests
that the light-cone picture cannot be valid point by point. The theoretical
prediction obtained in the light-cone expansion is not quite accurate and 
refers only to the smeared spectrum, not point by point. A related problem is 
the uncertainty in the charm quark mass. The charm quark mass is an important
source of uncertainty when one measures $\xi_c$, although, in principle, this
uncertainty can be eliminated experimentally by using the multivariate freedom
in $\xi_c$, as discussed previously. Since the spectrum 
$d\Gamma(B\to X_c\ell\nu)/d\xi_c$ has a sharp dependence on $\xi_c$, the
measurement precision required for the elimination of the charm quark mass 
uncertainty will be challenging experimentally.
 
The method proposed in this paper by measurements of the differential decay 
rate as a function of $\xi_u$ is, on the other hand, theoretically very 
reliable for a high precision $|V_{ub}|$ determination. The light-cone 
expansion works much better for the $B\to X_u\ell\nu$ decay because
a much larger momentum transfer with the maximum value of $q^2$ being $M^2$ 
can occur in the $B\to X_u\ell\nu$ decay than in the $B\to X_c\ell\nu$ decay. 
Many final hadronic states contribute to the $\xi_u$ spectrum above the charm 
threshold, without any preferential weighting towards the low-lying resonance 
states. Thus, we feel more confident in the method
of Eq.~(\ref{eq:vub}) for determining $|V_{ub}|$.
Applying the kinematic cut $\nu>M-M_D$, 
which leads to $\xi_u>1-M_D/M=0.65$ (see Fig.~1), 
one can discriminate between the inclusive $b\to u$ signal and $b\to c$ 
background. We find that about $99\%$ of the $b\to u$
events pass $\xi_u>1-M_D/M$, since the $\xi_u$ distribution peaks sharply 
at $\xi_u$ close to $1$, as demonstrated in Fig.~2. This 
discrimination between $b\to u$ and $b\to c$ events is even better than the 
method
of the hadronic invariant mass spectrum where about $90\%$ of the $b\to u$
events survive the cut on the hadronic invariant mass 
$M_X<M_D$ \cite{parton,barger,afalk,kim,bdu,jin3}. Thus, $b\to c$ background 
can be very efficiently suppressed and improved statistics can be achieved in 
the measurement of the spectrum 
$d\Gamma(B\to X_u\ell\nu)/d\xi_u$, permitting a precise determination 
of $|V_{ub}|$.  
We also anticipate that a reliable value for the inclusive charmless 
semileptonic branching fraction of the $B$ meson may be obtained through 
this measurement.

The residual hadronic uncertainty in $|V_{ub}|$ due to higher-order, 
power-suppressed corrections is expected to be at the level of $1\%$. The 
precision of this determination of $|V_{ub}|$ will mainly depend on its
experimental feasibility. The accuracy in $\nu$ and $q^2$, which determine
$\xi_u$, in the experiments seems essential for this method to work.
The experimental technique of neutrino 
reconstruction could well make this way of extracting $|V_{ub}|$ experimentally
feasible. If the neutrino can be reconstructed kinematically by inferring its
four-momentum from the missing energy and missing momentum in each event, then
it is possible to measure $\nu$ and $q^2$ and thus the scaling variable 
$\xi_u$. To reach $10\%$ 
precision in $|V_{ub}|$ requires, for instance, a measurement of $\nu$ and 
$\sqrt{q^2}$ with an error around $4\%$.
 
In conclusion, we have proposed new methods for determining $|V_{ub}|$ and
$|V_{cb}|$ by measuring the differential decay rates as a function of
$\xi_{u,c}$ in the inclusive semileptonic decays $B\to X_{u,c}\ell\nu$.
These methods are based on the light-cone expansion, which are, in principle,
model independent as it is in deep inelastic scattering.
We have shown that
the differential decay rates as a function of the scaling variables $\xi_{u,c}$
for the inclusive semileptonic $B$ meson decays $B\to X_{u,c}\ell\nu$ are 
proportional to the distribution function, which describes the leading
nonperturbative contribution, because of the light-cone dominance. 
This unique feature makes these observables ideally suitable for eliminating 
the dependence on the distribution 
function and thus avoiding the dominant hadronic uncertainties.  The 
integrated decay rates are, to a large extent, free of hadronic uncertainties 
by using the known normalization 
of the distribution function from current conservation, permitting precise 
determinations of $|V_{ub}|$ and $|V_{cb}|$, respectively. 
Moreover, the distribution function
cancels in the ratio of the differential decay rates at specific kinematic 
points, permitting a precise determination of $|V_{ub}/V_{cb}|$. The whole
methods are then free from model parameters. 
Since theoretical uncertainties associated with nonperturbative strong 
interactions are, to a large extent, avoided and perturbative 
corrections are calculable, these 
methods could yield very accurate values of $|V_{ub}|$ and $|V_{cb}|$. There
will be smaller residual theoretical uncertainties from higher-twist terms and
perturbative corrections.  

We wish to emphasize that the method by measurements of the spectrum 
$d\Gamma(B\to X_u\ell\nu)/d\xi_u$ is especially
promising for a high precision $|V_{ub}|$ determination. Theoretically,
this method is very reliable since the theoretical description based on the 
light-cone expansion works much better for $b\to u$ decays than $b\to c$ 
decays. 
With the dominant hadronic uncertainty being avoided, there is a residual
hadronic uncertainty in $|V_{ub}|$ only at the level of $1\%$. Experimentally,
it gives access to almost all $b\to u$ events, even more than the selection
against charm background based on the hadronic invariant mass spectrum. 
Therefore, at least potentially, this theoretically sound, clean and 
experimentally efficient method allows for a model-independent determination 
of $|V_{ub}|$ with a minimum overall (experimental and theoretical) error.

The decay spectra $d\Gamma(B\to X_u\ell\nu)/d\xi_u$ and
$d\Gamma(B\to X_c\ell\nu)/d\xi_c$ offer, on the other hand, the 
intrinsically most sensitive probe of long-distance strong interactions.
Measurements of the 
differential decay rates as a function of $\xi_{u,c}$ can be also used to 
extract the distribution function directly. This procedure is crucial for 
improving the theoretical accuracies on the determinations of $|V_{ub}|$
and $|V_{cb}|$ from the charged lepton energy spectra \cite{jin1,jp1}
and the hadronic invariant mass spectrum \cite{jin3} in inclusive semileptonic
$B$ meson decays.  A reliable value for the inclusive charmless semileptonic
branching fraction of the $B$ meson could also be obtained by measuring the 
spectrum $d\Gamma(B\to X_u\ell\nu)/d\xi_u$.

\acknowledgments
I would like to thank Emmanuel Paschos for discussions.
This work was supported in part by the Bundesministerium f\"ur Bildung, 
Wissenschaft, Forschung und Technologie, Bonn, FRG under grant 057DO93P(7), and
by the Australian Research Council.

{\tighten

} 

\end{document}